\PassOptionsToPackage{table}{xcolor}
\documentclass[10pt,twocolumn,letterpaper]{article}

\usepackage[pagenumbers]{cvpr}     
\usepackage[dvipsnames]{xcolor}

\definecolor{cvprblue}{rgb}{0.21,0.49,0.74}
\usepackage[pagebackref,breaklinks,colorlinks,allcolors=cvprblue]{hyperref}

\def\paperID{****} % *** Enter the Paper ID here
\def\confName{CVPR}
\def\confYear{2025}

\usepackage[table]{xcolor}
\usepackage{multirow}
\usepackage{graphicx}
\usepackage{booktabs}
\definecolor{cGreen}{HTML}{2e75b5}
\definecolor{cgray}{HTML}{FAFAFA}
\usepackage{hyperref}
\usepackage{url}
\usepackage{caption}
\usepackage{wrapfig}
\definecolor{orange}{HTML}{cc7700}
\definecolor{green}{HTML}{339955}
\definecolor{Highlight}{rgb}{0.12,0.49,0.85}
\definecolor{my_red}{HTML}{FE4444}
\usepackage{pifont}
\setcitestyle{circle}
\usepackage[capitalize]{cleveref}
\crefname{section}{Sec.}{Secs.}
\Crefname{section}{Section}{Sections}
\Crefname{table}{Table}{Tables}
\crefname{table}{Tab.}{Tabs.}
\newcommand{\best}[1]{{{\textcolor{red}{#1}}}}
\newcommand{\second}[1]{{\textcolor{blue}{{#1}}}}

\usepackage{xspace}
\newcommand{\NAME}{MambaIRv2\xspace}

\title{MambaIRv2: Attentive State Space Restoration}

\author{
  Hang Guo$^{1,*}$\enspace 
  Yong Guo$^{2,*}$\enspace 
  Yaohua Zha$^{1}$\enspace 
  Yulun Zhang$^{3}$\enspace 
  Wenbo Li$^{4}$ \\
  Tao Dai$^{5,\dagger}$\enspace 
  Shu-Tao Xia$^{1,6}$\enspace 
  Yawei Li$^{7}$ \\
  \textsuperscript{1}Tsinghua University\enspace 
  \textsuperscript{2}Max Planck Institute for Informatics\\ 
  \textsuperscript{3}Shanghai Jiao Tong University\enspace
  \textsuperscript{4}The Chinese University of Hong Kong\\
  \textsuperscript{5}Shenzhen University\enspace
  \textsuperscript{6}Peng Cheng Laboratory\enspace
  \textsuperscript{7}ETH Z\"{u}rich \\
  \vspace{-6mm}
}

\begin{document}

\renewcommand{\thefootnote}{\fnsymbol{footnote}}

\maketitle
\footnotetext[1]{Equal contribution}
\footnotetext[2]{Corresponding author}

\begin{abstract}
The Mamba-based image restoration backbones have recently demonstrated significant potential in balancing global reception and computational efficiency. However, the inherent causal modeling limitation of Mamba, where each token depends solely on its predecessors in the scanned sequence, restricts the full utilization of pixels across the image and thus presents new challenges in image restoration. In this work, we propose MambaIRv2, which equips Mamba with the non-causal modeling ability similar to ViTs to reach the attentive state space restoration model. Specifically, the proposed attentive state-space equation allows to attend beyond the scanned sequence and facilitate image unfolding with just one single scan. Moreover, we further introduce a semantic-guided neighboring mechanism to encourage interaction between distant but similar pixels. Extensive experiments show our MambaIRv2 outperforms SRFormer by \textbf{even 0.35dB} PSNR for lightweight SR even with \textbf{9.3\% less} parameters and suppresses HAT on classic SR by \textbf{up to 0.29dB}. Code is available at \url{https://github.com/csguoh/MambaIR}.
\end{abstract}

\vspace{-6mm}

\section{Introduction}

Image restoration aims to recover high-quality images from low-quality observations, tackling various sub-problems such as image super-resolution, image denoising, and JPEG compression reduction, and others. With the advent of deep learning, state-of-the-art performance has been consistently achieved. Early works primarily utilized convolutional neural networks (CNNs) as the backbone~\cite{DnCNN,dong2014learning,lim2017enhanced,zhang2018residual,dai2019second}. Later, vision transformers (ViTs)~\cite{dosovitskiy2020image} gained popularity for their superior performance~\cite{chen2021pre,liang2021swinir,chen2023activating,li2023grl,chen2023dual}. More recently, the selective state-space model (Mamba)~\cite{gu2023mamba} has been explored, showing considerable potential as an alternative backbone for image restoration tasks~\cite{guo2024mambair,shi2024vmambair}.

\begin{figure}[!t]
    \centering
    \includegraphics[width=0.95\linewidth]{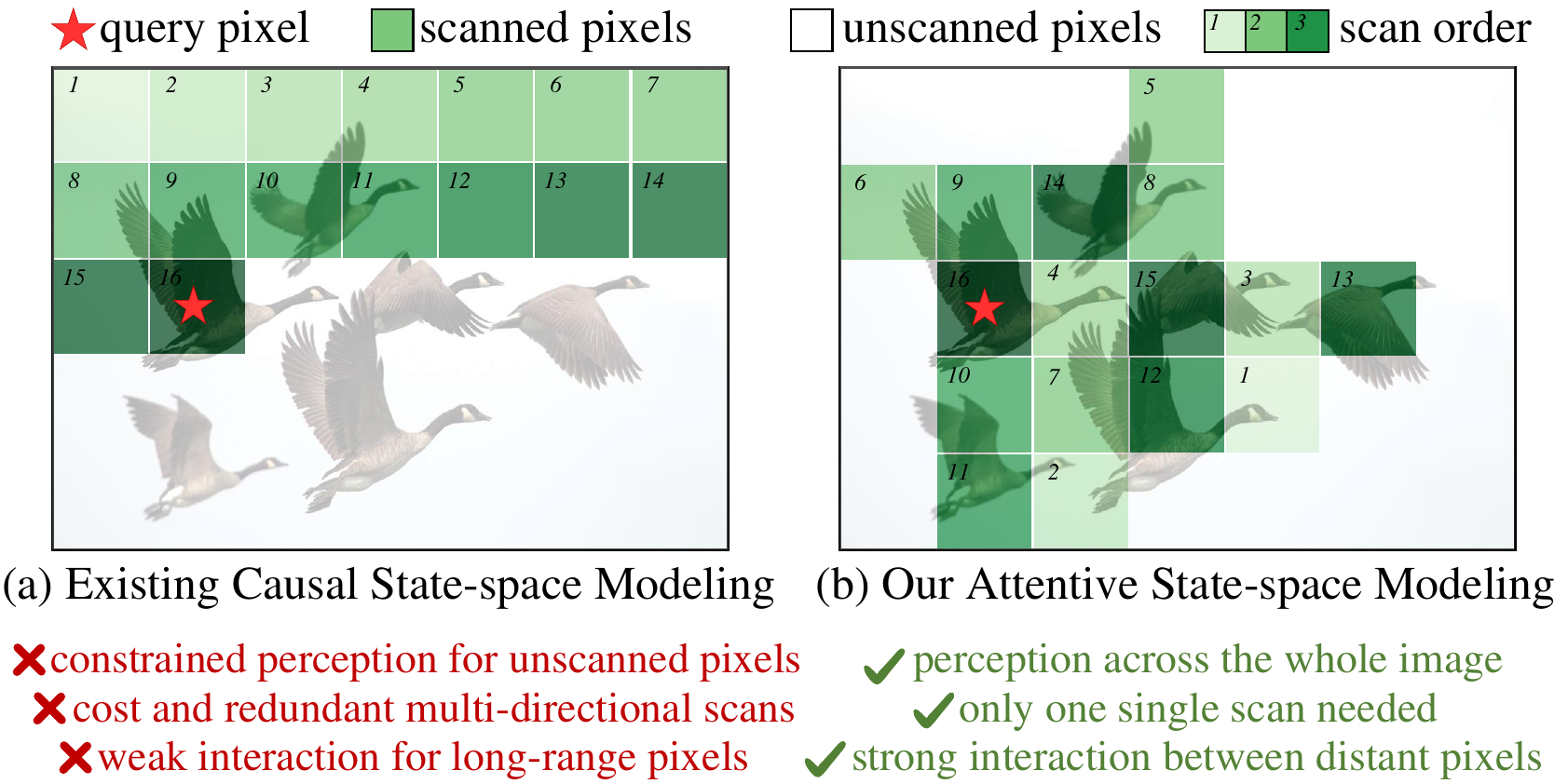}
    \vspace{-2mm}
    \caption{ (a) The existing method~\cite{guo2024mambair} suffers from the adverse effects of the causal nature of Mamba (the multi-directional scans are not shown for presentation clarity). (b) The proposed MambaIRv2 can achieve attentive state-space modeling that embeds ViT-like non-causal properties into Mamba.}
    \label{fig:motivation}
    \vspace{-6mm}
\end{figure}

Despite its potential, existing Mamba-based methods face significant challenges, particularly due to their reliance on causal state-space modeling. 
Specifically, existing methods~\cite{guo2024mambair} unfold the 2D image with a predefined scanning rule to generate the 1D token sequence. However, in Mamba, each pixel is modeled based solely on its preceding pixels in the scanned sequence, \textit{i.e.}, the causal property, which results in several detrimental effects for non-causal image restoration tasks.
\textbf{First}, as shown in~\cref{fig:motivation}(a), the query pixel can only capture information from its preceding ones and cannot perceive subsequent pixels, which results in under-utilization of helpful pixels across the image. 
\textbf{Second}, the inherent causal property leads to the necessity of multi-directional scans, which is widely adopted by existing approaches~\cite{guo2024mambair,shi2024vmambair,weng2024mamballie} for mitigating information loss. Yet, this multi-scanning inevitably increases the computational complexity, particularly for high-resolution inputs. Furthermore, an empirical investigation in~\cref{sec:preliminary} reveals that there is also substantial information redundancy among these multi-directional scans. 
\textbf{Third}, our findings in~\cref{sec:preliminary} demonstrate that Mamba~\cite{gu2023mamba} is prone to long-range decay in token interaction, meaning distant tokens in the sequence have diminished interactions. Consequently, even previously scanned pixels that are distant yet relevant cannot be effectively utilized by the query pixel.

In this work, we propose MambaIRv2, aiming to address the adverse effect of causal state-space modeling. Since the ViTs~\cite{vaswani2017attention,dosovitskiy2020image} naturally support non-causal processing, our key idea is to integrate ViT-like non-causal modeling into the Mamba-based methods. To this end, we begin by delving deep into the connection between attention and state space for valuable insights. Our in-depth analysis in \cref{sec:connect-attn-ssm} reveals that the output matrix of the state-space equation resembles the query in the attention mechanism. This similarity inspires us to utilize the output matrix to \textbf{``query''} relevant pixels in the unscanned sequence. 
Benefiting from attending beyond the scanned sequences, this strategy also naturally eliminates the need for multi-directional scanning. Moreover, to encourage the interaction between distant but relevant pixels, we propose to restructure the image to place similar pixels spatially closer within the 1D sequence. In this way, it allows for semantic rather than spatial sequence modeling, thus mitigating the impact of long-range decay. Since the proposed method allows the Mamba to behave similarly to the attention, we thus refer to it as \textbf{``attentive state-space restoration''}.

Overall, we make three key contributions: 
\textbf{I.} We propose the \textbf{Attentive State-space Equation (ASE)}, which leverages the prompt learning~\cite{jia2022visual} within the original state space equation of Mamba to query semantically similar pixels beyond the scanned sequences. In detail, the prompts are designed to represent sets of pixels that are similar across the entire image, and we then incorporate the representative prompts into the output matrix of the state-space equation through residual addition to derive our ASE. As the core component, the proposed ASE not only alleviates the causal nature of Mamba for improved performance but also enables single-pass scanning for boosted efficiency.
\textbf{II.} We further develop the \textbf{Semantic Guided Neighboring (SGN)} to encourage strong interaction between distant yet similar pixels. Specifically, we first assign the corresponding semantic label to each pixel. Then we restructure the image based on these labels to generate the semantic-neighboring 1D sequence, where semantically similar pixels are also spatially close to each other. Thanks to the mitigation of the long-range decay of Mamba, SGN facilitates effective interaction between pixels that are distant in the original image.
\textbf{III.} Integrating the two core modules and other auxiliary parts, we present MambaIRv2, an attentive state-space restoration method that equips Mamba’s state-space modeling with ViT-like non-causal capabilities. Extensive experiments demonstrate that MambaIRv2 significantly improves \textbf{both effectiveness and efficiency}. In particular, MmabaIRv2 outperforms state-of-the-art Transformer-based baseline SRFormer~\cite{zhou2023srformer} by 0.35dB on Urban100 dataset for $2\times$ lightweight SR with 9.3\% less parameters, and HAT~\cite{chen2023activating} by 0.29dB for $2\times$ classical SR on the Manga109 dataset.

\section{Related Work}
Recent years have witnessed great advancements in the domain of image restoration~\cite{ali2023vitsurvy,jiang2024allinonesurvy,he2025diffusionsurvy}. Early attempts usually adopt the convolutional neural networks (CNNs), such as SRCNN~\cite{dong2014learning} for image super-resolution, DnCNN~\cite{DnCNN} for image denoising, and ARCNN~\cite{dong2015compression} for JPEG compression artifacts reduction. To further enhance the performance of CNN-based methods, various techniques have been introduced. For instance, EDSR~\cite{kim2016accurate} employ the residual connection strategy to allow the training of very deep neural networks, RDN~\cite{zhang2018residual} uses the dense connection to improve model representation ability. RCAN~\cite{zhang2018image} introduces the channel attention for selecting salient channels, followed by SAN~\cite{dai2019second} which uses the second-order attention for performance improvement. Despite the great progress of CNN-based methods, the convolution operator inherently restricts the receptive field to the local kernel, preventing interaction between distant pixels.

As Transformer~\cite{vaswani2017attention} has proven its effectiveness in multiple computer vision tasks, applying Transformer for image restoration thus appears to be promising. However, the direct application of vanilla self-attention, which exhibits quadratic computational complexity with the input size, is costly and impractical. To improve the efficiency of attention, a variety of techniques have been developed. For example, IPT~\cite{chen2021pre} divides one image into several small patches and processes each patch independently with self-attention. After that, SwinIR~\cite{liang2021swinir} further introduces the shifted window self-attention~\cite{liu2021swin} to improve the performance. ART~\cite{zhang2023accurate} and OminiSR~\cite{wang2023omni} utilize sparse attention to expand the receptive field by enlarging the attention windows. GRL~\cite{li2023grl} adopts the anchor attention to learn the local, regional, and global image hierarchies. Recently, the ATD~\cite{zhang2024transcending} uses the adaptive token dictionary to store input-agnostic knowledge, thus allowing the attention to attend information out of the local window.

To balance the efficient computation and global receptive fields, the Mamba~\cite{gu2023mamba} has recently been explored in image restoration with promising results. MambaIR~\cite{guo2024mambair} is among the first to introduce Mamba for image restoration and addresses two specific challenges, \textit{i.e.}, local pixel forgetting and channel redundancy. Since then, the Mamba model has been explored in various image restoration tasks. FreqMamba~\cite{zhen2024freqmamba} uses the state space model in the Fourier domain for image deraining to perceive global degradation. MambaLLIE~\cite{weng2024mamballie} improves the state space equation to allow the locality enhancement for low-light image enhancement tasks.
Moreover, Mamba has also achieved promising results in image dehazing~\cite{zheng2024mambadehaze}, debluring~\cite{gao2024mambadebulr}, and other tasks~\cite{zhang2024vmambasci,wu2024rainmamba,lin2024pixmamba,xie2024fusionmamba,qin2024mambavc,bai2024retinexmamba,qiao2024himamba}. However, existing methods still struggle with the causal modeling nature of Mamba. Given image restoration as a non-causal task, this mismatch can lead to limited performance as well as inefficiency.

\section{Motivation}
\label{sec:preliminary}
\noindent \textbf{Mamba-based Image Restoration.}
The existing state-space restoration methods are mainly developed from the Mamba~\cite{gu2023mamba} architecture. Formally, the Mamba adopts the discrete state space equation to model the interaction among tokens:
\begin{equation}
\begin{aligned}
\label{eq:ssm}
&h_{i}=\mathbf{\overline{A}} h_{i-1}+\mathbf{\overline{{B}}} x_i,\\
&y_{i}=\mathbf{C} h_i+ \mathbf{D} x_i,
\end{aligned}
\end{equation}
where the $\mathbf{\overline{A}}=\exp(\mathbf{\Delta}\mathbf{A})$  is the control matrix, the $\mathbf{\overline{B}} = (\mathbf{\Delta}\mathbf{A})^{-1}(\exp(\mathbf{\Delta}\mathbf{A})-\mathbf{I})\mathbf{\Delta {B}} \approx \mathbf{\Delta} \mathbf{B}$ is the input matrix, and the $\mathbf{C}$ is the output matrix. \cref{eq:ssm} indicates that the $i$-th token completely depends on its previous $i-1$ tokens, \textit{i.e.}, the state-space modeling possesses causal properties. Although this causal nature is helpful for autoregressive tasks like NLP, it poses challenges for image restoration.

\begin{figure}[!t]
    \centering
    \includegraphics[width=0.96\linewidth]{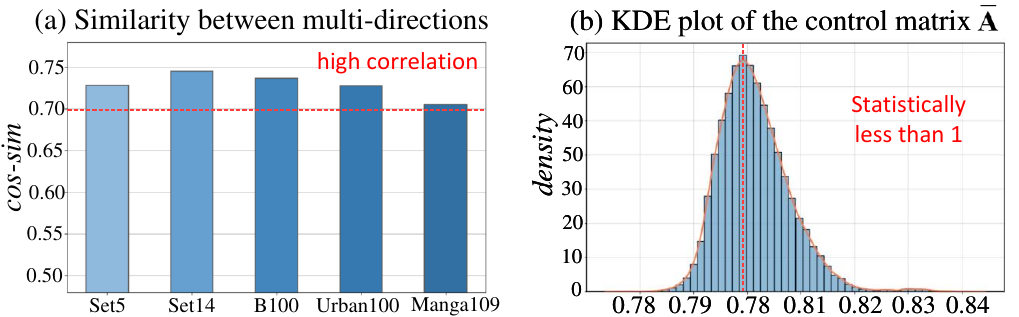}
    \vspace{-2mm}
    \caption{(a) We compute the cosine similarity of scanned features across all 4 directions and all layers in MambaIR~\cite{guo2024mambair}. (b) The kernel density estimation of the distribution of the control matrix in MambaIR~\cite{guo2024mambair}.}
    \vspace{-3mm}
    \label{fig:preliminary}
\end{figure}

\noindent \textbf{Challenges from Causal Modeling.}
Existing Mamba-based methods usually adopt a specific scanning strategy to unfold the 2D image into a 1D sequence for sequential modeling with Mamba. However, the $i$-the pixel can only see constrained $i-1$ pixels of the entire image, failing to globally utilize similar pixels. To this end, current methods typically employ multi-directional scans to allow for a broader receptive field, which is inevitably accompanied by an increase in computational complexity. Furthermore, as shown in~\cref{fig:preliminary}(a), the similarity of different scanned sequences on all testing datasets reaches even above 0.7, indicating a high correlation with large redundancy. Moreover, the Mamba~\cite{gu2023mamba} itself also possesses long-range decay defects due to its causal nature. Specifically, the interaction between pixels can be quantitatively denoted by the power of control matrix $\overline{\textbf{A}}^k$, where $k$ is the pairwise distance (the proof is given in the \textit{Suppl.}). In~\cref{fig:preliminary}(b), we show that the learned $\overline{\textbf{{A}}}$ is statistically less than 1. As a result, the interaction $\overline{\textbf{A}}^{k}$ will become weak when two pixels are far apart, \textit{i.e.}, a large $k$, indicating current methods fail to utilize distant but useful scanned pixels.

\begin{figure*}[!t]
    \centering
    \includegraphics[width=0.96\linewidth]{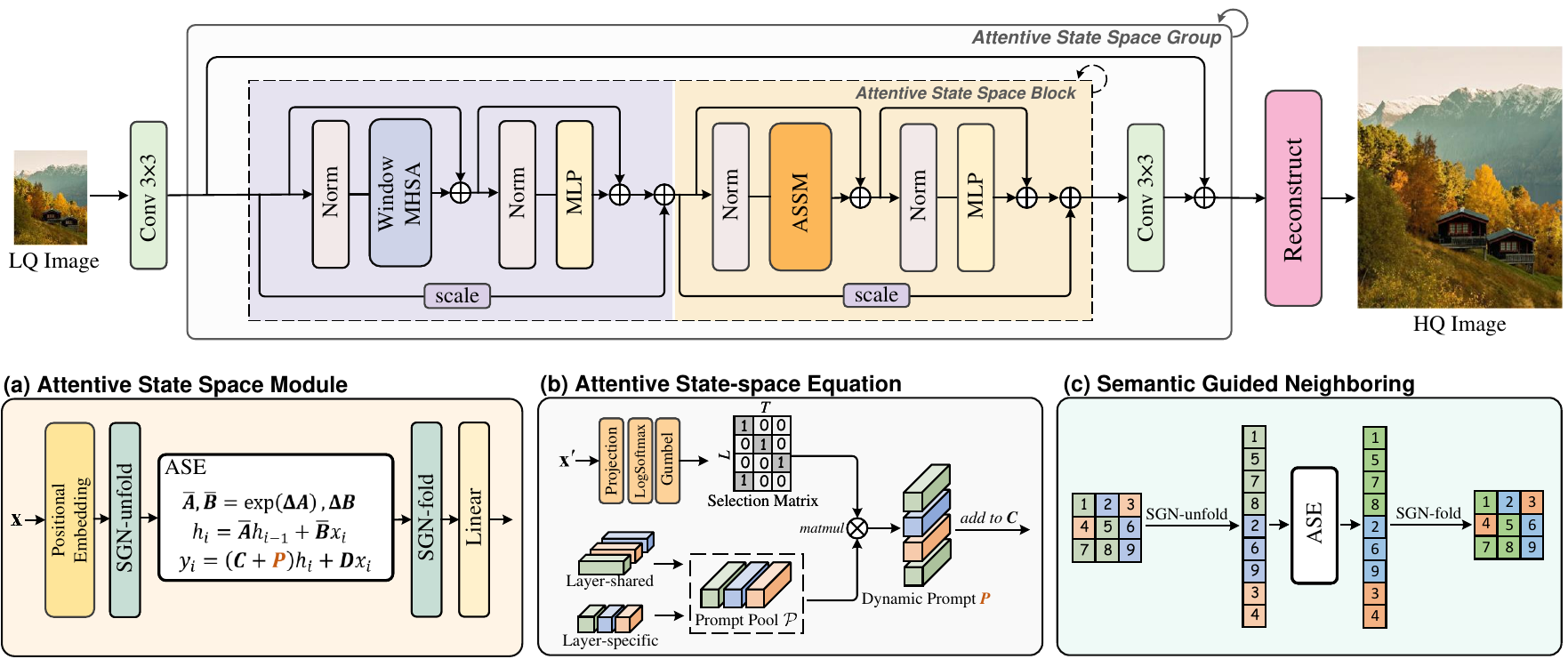}
    \vspace{-1mm}
    \caption{The overall architecture of our proposed MambaIRv2, as well as the (a) Attentive State Space Module (ASSM), (b) Attentive State-space Equition (ASE), and (c) Semantic Guided Neighboring (SGN). }
    \vspace{0mm}
    \label{fig:pipeline}
\end{figure*}

\section{Attentive State Space Restoration}
In the following, we aim to address the causal nature of Mamba through the proposed attentive state space restoration. To get started, we visit the mathematical connection between state space and attention in~\cref{sec:connect-attn-ssm} for insights of subsequent design. Then, we detail the specific techniques of our attentive state space module in~\cref{sec:attntive-state-space-module}. In~\cref{sec:hierarchy-learning}, we give the overall architecture of the proposed methods.

\subsection{Bridging Attention and State-Space}
\label{sec:connect-attn-ssm}
As pointed out by~\cite{han2024demystify}, the state space has a strong relationship to attention, which may potentially offer insights to incorporate non-causal modeling ability into Mamba. In this section, we first reformulate attention and state space into the common form for comparison, followed by a detailed connection analysis.

\noindent \textbf{Reformulation of Attention.} Since the Mamba belongs to causal models with linear complexity, we adopt the corresponding causal linear attention~\cite{katharopoulos2020linearattn} as its counterpart. Specifically, given the query, key and value matrix $\mathbf{Q}$, $\mathbf{K}$, $\mathbf{V}$, the output of linear attention is computed as follows:
\begin{equation}
y_{i}=\sum_{j=1}^{i}\frac{\mathbf{Q}_{i}\mathbf{K}_{j}^{\top}}{\sum_{t=1}^{i}\mathbf{Q}_{i}\mathbf{K}_{t}^{\top}}\mathbf{V}_{j}=\frac{\mathbf{Q}_{i}\left(\sum_{j=1}^{i}\mathbf{K}_{j}^{\top}\mathbf{V}_{j}\right)}{\mathbf{Q}_{i}\left(\sum_{t=1}^{i}\mathbf{K}_{t}^{\top}\right)}.
\end{equation}
Denote the $\mathbf{S}_i=\sum_{j=1}^i \mathbf{K}_j^\top \mathbf{V}_j, \mathbf{Z}_i=\sum_{t=1}^i \mathbf{K}_t^\top $, then the formulation of the linear attention can be rewritten as:
\begin{equation}
\label{eq:attn}
y_i=\mathbf{Q}_i \mathbf{S}_i/\mathbf{Q}_i \mathbf{Z}_i,
\end{equation}
where  $\mathbf{S}_i=\mathbf{S}_{i-1} + \mathbf{K}_i^\top \mathbf{V}_i$, and $\mathbf{Z}_i=\mathbf{Z}_{i-1}+\mathbf{K}_i^\top$. To allow subsequent connection analysis, we further reformulate ~\cref{eq:attn} to the common form as follows:
\begin{equation}
\begin{aligned}
\label{eq:compare-attn}
&\mathbf{S}_i= \mathbf{I} \mathbf{S}_{i-1} + \mathbf{K}_i^\top \mathbf{V}_i,\\
&y_i=\mathbf{Q}_i \mathbf{S}_i / \mathbf{Q}_i \mathbf{Z}_i + \mathbf{O} x_i,
\end{aligned}
\end{equation}
where the $\mathbf{I}$ and $\mathbf{O}$ denotes identity and zero matrix, respectively. $x_i$ is the input token at the $i$-th step.

\noindent \textbf{Reformulation of State Space.}
Starting with state-space equation in~\cref{eq:ssm}, note that $\mathbf{\overline{B}} x_i \approx \mathbf{\Delta} \mathbf{B} x_i=\mathbf{B} (\mathbf{\Delta} x_i)$, ~\cref{eq:ssm} can then be reformulated to the common form as:
\begin{equation}
\begin{aligned}
\label{eq:compare-ssm}
&{h}_i=\mathbf{\overline{A}} {h}_{i-1}+\mathbf{B}(\mathbf{\Delta} {x}_i), \\
&y_i=\mathbf{C} h_i/\mathbf{I}+\mathbf{D} {x}_i.
\end{aligned}\end{equation}

\noindent \textbf{Connection Analysis.}
By comparing~\cref{eq:compare-attn} and ~\cref{eq:compare-ssm}, it can be seen there is a close mathematical similarity between attention and state space, \textit{i.e.}, $h_i \sim \mathbf{S}_i$, $\mathbf{B} \sim \mathbf{K}^\top$, and $\mathbf{C} \sim \mathbf{Q}$. It should be noted that the $\mathbf{Q}_i \mathbf{Z}_i$ in ~\cref{eq:compare-attn} is to ensure the attention score is normalized with summation 1 and can be approximately ignored if we relax the normalization constraints. Therefore, the above observation motivates us to delve deep into the output matrix $\mathbf{C}$ in~\cref{eq:ssm}, which plays a role similar to the query in the attention mechanism. The core idea is to integrate the information of the unscanned sequence into $\mathbf{C}$, thus allowing $\mathbf{C}$ to attentively ``query'' unseen pixels to facilitate the restoration of $x_i$.

\subsection{Attentive State Space Module}
\label{sec:attntive-state-space-module}

In this section, we introduce the attentive state-space module (ASSM), which acts as the core block of our MambaIRv2 to enable non-causal modeling with Mamba. As shown in~\cref{fig:pipeline}(a), given the input feature $\mathbf{x} \in \mathbb{R}^{H\times W \times  C}$, where $H$ and $W$ are the height and width, respectively, and $C$ is the channel dimension, we first apply the positional encoding~\cite{chu2021conditional} on $\mathbf{x}$ to preserve the original structure information. After that, we propose the Semantic Guided Neighboring (SGN) to unfold the 2D image into 1D sequences for subsequent Attentive State-space Equation (ASE) modeling. Finally, another SGN is employed as the inverse operator of the previous one to fold the sequence back to the image followed by a linear projection to obtain the block output. More details are given below.

\noindent \textbf{Attentive State-space Equation.}
As analyzed in~\cref{sec:connect-attn-ssm}, we aim to modify the output matrix $\mathbf{C} \in \mathbb{R}^{L\times d}$, where $L=HW$ is the flattened image sequence length and $d$ is the number of hidden states in Mamba, to globally query related pixels across the image. To this end, we propose the Attentive State-space Equation (ASE) which develops from the original state-space equation of Mamba but possesses a non-causal nature. As shown in~\cref{fig:pipeline}(b), the proposed ASE incorporates prompts, which learn to represent a certain set of pixels with similar semantics, into $\mathbf{C}$ to supplement the missing information of the unseen pixels. Specifically, we first construct the prompt pool $\mathcal{P}\in \mathbb{R}^{T \times d}$, where $T$ is the number of prompts in $\mathcal{P}$. For parameterization of $\mathcal{P}$, we employ the semantic decoupling for better interpretability:
\begin{equation}
\mathcal{P} = \mathbf{M} \mathbf{N}, \quad \mathbf{M} \in \mathbb{R}^{T\times r}, \quad \mathbf{N} \in \mathbb{R}^{r \times d},   
\end{equation}
where $\mathbf{N}$ is shared across different blocks, $\mathbf{M}$ is block-specific, and $r$ is the inner rank with $r \ll \min\{T, d\}$. The main idea behind the semantic decoupling is that we want different blocks to share similar feature space, $\textit{i.e.}$, $\mathbf{N}$ is shared, while the combination coefficients of the shared features can be varying for different blocks $\textit{i.e.}$, $\mathbf{M}$ is specific.

After that, we develop routing strategies to select from $\mathcal{P}$ to obtain $L$ instance-specific prompts $\mathbf{P} \in \mathbb{R}^{L \times d}$, which will be added into $\mathbf{C}$ to include information of unscanned pixels. In detail, given the flattened input feature $\mathbf{x'} \in \mathbb{R}^{L \times C}$, we employ a linear layer to project the channel dimension of $\mathbf{x'}$ from $C$ to $T$, followed by the $\mathrm{LogSoftmax}$ to predict the log probability, which indicates the probability of each prompt in $\mathcal{P}$ being sampled by $\mathbf{x}'_i, i=1,2,\cdots L$. After that, we introduce the gumbel-softmax~\cite{jang2016gumbel} trick on the log probability to allow differentiable prompt selection to obtain the one-hot routing matrix $\mathbf{R} \in \mathbb{R}^{L \times T}$. Then, the instance-specific prompt $\mathbf{P} \in \mathbb{R}^{L\times d}$ is generated through the matrix multiplication as $\mathbf{P} = \mathbf{R} \mathcal{P}$. Finally, we incorporate $\mathbf{P}$ into $\mathbf{C}$ through residual addition to formulate our attentive state-space equation:
\begin{equation}
\begin{aligned}
\label{eq:ase-ssm}
&h_{i}=\mathbf{\overline{A}} h_{i-1}+\mathbf{\overline{{B}}} x_i,\\
&y_{i}=(\mathbf{C + \textcolor[RGB]{197, 90, 17}{P}}) h_i+ \mathbf{D} x_i.
\end{aligned}
\end{equation}
The learned prompts allow for the attention-like capability to query pixels across the whole image, and we also present a visualization on the attentive map in~\cref{fig:attntive-map}. By injecting the prompts that represent the set of similar pixels, the proposed ASE can effectively alleviate constrained perception for unscanned pixels. As another advantage, it allows scanning with only one single direction, eliminating the high computational cost and redundancy of multi-directional scans in existing methods.

\noindent \textbf{Semantic Guided Neighboring.}
As pointed out in~\cref{sec:preliminary}, the causal modeling property of Mamba leads to the detrimental effects of long-range decay. In existing Mamba-based image restoration methods, pixels that are distant in the original image are usually still far apart in the unfolded sequence, causing the weak utilization of the query pixel for the already scanned pixels which are spatially distant but similar. To this end, we propose the Sematic Guided Neighboring (SGN) as shown in~\cref{fig:pipeline}(c). Our key insight is that different from the autoregressive language modeling, the image restoration is a non-causal task and all pixels are observable at once, therefore we can re-define the token neighborhood to enable semantically similar tokens to be spatially closer in the unfolded sequence. Following this idea, we first determine the semantic label of each pixel. Note that the routing matrix $\mathbf{R}$ in the ASE, which has learned the prompt category of each pixel, we thus employ this off-the-shelf semantics to restructure the image. Specifically, we propose the SGN-unfold which groups pixels with the $i$-th prompt category together to form the $i$-th semantic group and then combines different groups according to the category value $i$ to generate the semantic-neighbored sequence. After that, we feed this sequence into the proposed ASE for state-space modeling. At last, we employ the SGN-fold which performs as the inverse transformation of SGN-unfold to reshape the semantic-space sequence back to the spatial-space feature map to obtain the output.

\subsection{Overall Network Architecture}
\label{sec:hierarchy-learning}
The proposed ASSM can efficiently capture the global dependencies using the Mamba model. We then further consider modeling local interactions, which have been shown crucial for Mamba-based approaches~\cite{guo2024mambair,weng2024mamballie}. Since the scanning only once in the ASSM provides more parameter budgets, we thus opt for the powerful window multi-head self-attention (MHSA)~\cite{liu2021swin} to enhance local interactions within the window, which together with the ASSM compose the basic elements of our MambaIRv2.
As shown in~\cref{fig:pipeline}, given a low-quality image as the input, we first utilize the $3\times3$ convolution layer to extract shallow features. Then the shallow features are fed into several Attentive State Space Groups (ASSGs), where each 
groups contain multiple Attentive State Space Blocks (ASSBs).
For each ASSB, we consider progressive local-to-global modeling to form the image hierarchy~\cite{li2023grl}. We use Norm and Token Mixer, followed by Norm and FFN to form the template, and employ window MHSA and ASSM as the instantiations for the Token Mixer of the local and global parts, respectively. In addition, two residual connections with learnable scales are introduced~\cite{guo2024mambair,chen2023recursive}. After the ASSGs, we utilize the task-specific reconstruction modules, \textit{e.g.}, pixelshuffle for super-resolution, and convolution for denoising, to obtain the high-quality image output.

\section{Experiments}

Following previous image restoration works~\cite{liang2021swinir,guo2024mambair}, we conduct experiments on three representative image restoration tasks, \textit{i.e.}, image super-resolution including classic SR and lightweight SR, JPEG compression artifact reduction (JPEG CAR), and Gaussian color image denoising.

\subsection{Experimental Settings}

% \noindent \textbf{Training Details.} 
In accordance with previous works, we perform data augmentation by applying horizontal flips and random rotations of $90^\circ, 180^\circ$, and $270^\circ$. Additionally, we crop the original images into $64 \times 64$ patches for image SR and $128 \times 128$ patches for image denoising during training. For image SR, we use the pre-trained weights from the $2\times$ model to initialize those of $3\times$ and $4\times$ and halve the learning rate and total training iterations to reduce training time~\cite{lim2017enhanced}. To ensure a fair comparison, we adjust the training batch size to 32 for image SR and 8 for image denoising and JPEG CAR. We employ the Adam~\cite{kingma2014adam} as the optimizer for training our MambaIRv2 with $\beta_1 = 0.9, \beta_2 = 0.999$. Similar to previous training protocol~\cite{liang2021swinir}, we use the $L_1$ loss for image SR, and Charbonnier loss~\cite{charbonnier1994two} for denoising and JPEG CAR. The initial learning rate is set at $2 \times 10^{-4}$ and is halved when the training iteration reaches specific milestones. For classic image SR, we provide three variants with different parameters including the small, base, and large versions (MambaIRv2-S, MambaIRv2-B, MambaIRv2-L). Due to the page limit, more details are provided in the \textit{Suppl.}

\subsection{Ablation Study}

We conduct ablations with MambaIRv2-light $2\times$ SR model trained for 250K iterations on the DIV2K dataset.

\begin{table}[!tb]
\centering
\caption{Ablation on the effectiveness of different components.}
\label{tab:ablation-components}
\vspace{-3mm}
\setlength{\tabcolsep}{8pt}
\scalebox{0.75}{
\begin{tabular}{@{}ccc|cc|cc@{}}
\toprule
\multirow{2}{*}{MHSA} & \multirow{2}{*}{ASE} & \multirow{2}{*}{SGN} & \multicolumn{2}{c|}{\textbf{Urban100}} & \multicolumn{2}{c}{\textbf{Manga109}} \\
          &           &           & PSNR & SSIM & PSNR & SSIM \\ \midrule
\ding{52} &       &        & 32.89&	0.9343&39.11&0.9772     \\
\ding{52} & \ding{52} &  &  32.94&	0.9351&39.20&0.9780     \\
\rowcolor[HTML]{EFEFEF} 
\ding{52} & \ding{52} & \ding{52} &    \textbf{32.97} &	\textbf{0.9355} & 	\textbf{39.24}	& \textbf{0.9784}   \\ \bottomrule
\end{tabular}%
}
\vspace{-2mm}
\end{table}

\begin{table}[!tb]
\centering
\caption{Ablation experiments on different injection positions of the learnable prompts in the ASE.}
\label{tab:ablation-prompt-ASE}
\vspace{-3mm}
\setlength{\tabcolsep}{7.2pt}
\scalebox{0.73}{
\begin{tabular}{@{}l|cc|cc|cc@{}}
\toprule
\multirow{2}{*}{positions} & \multicolumn{2}{c|}{\textbf{Set14}} & \multicolumn{2}{c|}{\textbf{Urban100}} & \multicolumn{2}{c}{\textbf{Manga109}} \\
             & PSNR & SSIM & PSNR & SSIM & PSNR & SSIM \\ \midrule
$\mathbf{B}$ &  33.97	& 0.9215&  32.96&	\textbf{0.9356} &	39.23&	0.9781   \\
$\Delta$     & 33.92&	0.9211&   32.93&0.9350&	39.19&0.9779      \\
y            & 33.97&	0.9210& 32.94&0.9351&39.21&	0.9782      \\
\rowcolor[HTML]{EFEFEF} 
$\mathbf{C}$ &   \textbf{33.99}	& \textbf{0.9216} &  \textbf{32.97} &	0.9355 & 	\textbf{39.24}	& \textbf{0.9784}    \\ \bottomrule
\end{tabular}%
}
\vspace{-3mm}
\end{table}

\begin{table*}[!t]
\centering
\caption{Quantitative comparison on \textit{\textbf{lightweight image super-resolution}} with state-of-the-art methods. The best and the second best results are in \best{red} and \second{blue}.}
\label{tab:lightSR}
\vspace{-3mm}
\setlength{\tabcolsep}{8pt}
\scalebox{0.75}{
\begin{tabular}{@{}l|c|c|c|cc|cc|cc|cc|cc@{}}
\toprule
 & & & & \multicolumn{2}{c|}{\textbf{Set5}} &
  \multicolumn{2}{c|}{\textbf{Set14}} &
  \multicolumn{2}{c|}{\textbf{BSDS100}} &
  \multicolumn{2}{c|}{\textbf{Urban100}} &
  \multicolumn{2}{c}{\textbf{Manga109}} \\
\multirow{-2}{*}{Method} & \multirow{-2}{*}{scale}& \multirow{-2}{*}{\#param}& \multirow{-2}{*}{MACs} & PSNR  & SSIM   & PSNR  & SSIM   & PSNR  & SSIM   & PSNR  & SSIM   & PSNR  & SSIM   \\ \midrule
CARN~\cite{ahn2018fast} & $2\times$  & 1,592K & 222.8G
& 37.76
& 0.9590
& 33.52
& 0.9166
& 32.09
& 0.8978
& 31.92
& 0.9256
& 38.36
& 0.9765
\\
LatticeNet~\cite{luo2020latticenet} & $2\times$  & 756K & 169.5G
& {38.13}
& 0.9610
& {33.78}
& {0.9193}
& {32.25}
& {0.9005}
& {32.43}
& {0.9302}
& -
& -
\\
SwinIR-light~\cite{liang2021swinir}& $2\times$  & 910K & {244.2G} 
& 38.14
& {0.9611}
& {33.86}
& {0.9206}
& {32.31}
& {0.9012}
& {32.76}
& {0.9340}
& {39.12}
& {0.9783}
\\
MambaIR-light~\cite{guo2024mambair} & $2\times$  & 905K & 334.2G
& {38.13}
& {0.9610}
& {33.95}
& {0.9208}
& {32.31}
& {0.9013}
& {32.85}
& {0.9349}
& {39.20}
& {0.9782}
\\
ELAN~\cite{zhang2022elan} & $2\times$  & 621K & 203.1G
& 38.17 
&0.9611 
&33.94
&0.9207 
&32.30 
&0.9012 
&32.76 
&0.9340
&39.11
&0.9782
\\
SRFormer-light~\cite{zhou2023srformer} & $2\times$  & 853K & 236.3G
&\second{38.23}
&\second{0.9613}
&\second{33.94}
&\second{0.9209}
&\second{32.36}
&\second{0.9019}
&\second{32.91}
&\second{0.9353}
&\second{39.28}
&\second{0.9785}
\\
\rowcolor[HTML]{EFEFEF} 
MambaIRv2-light & $2\times$  & 774K & 286.3G 
&\best{38.26}
&\best{0.9615}
&\best{34.09}
&\best{0.9221}
&\best{32.36}
&\best{0.9019}
&\best{33.26}	
&\best{0.9378}	
&\best{39.35}	
&\best{0.9785}
\\
\midrule
CARN~\cite{ahn2018fast} & $3\times$  & 1,592K  & 118.8G
& 34.29
& 0.9255
& 30.29
& 0.8407
& 29.06
& 0.8034
& 28.06
& 0.8493
& 33.50 
& 0.9440
\\ 
LatticeNet~\cite{luo2020latticenet} & $3\times$  & 765K & 76.3G 
& {34.53}
& {0.9281}
& {30.39}
& {0.8424}
& {29.15}
& {0.8059}
& {28.33}
& {0.8538}
& -
& -
\\
SwinIR-light~\cite{liang2021swinir} & $3\times$  & 918K & {111.2G} 
& {34.62}
& {0.9289}
& {30.54}
& {0.8463}
& {29.20}
& {0.8082}
& {28.66}
& {0.8624}
& {33.98}
& {0.9478}
\\ 
MambaIR-light & $3\times$  & 913K & {148.5G} %
& {34.63}
& {0.9288}
& {30.54}
& {0.8459}
& {29.23}
& {0.8084}
& {28.70}
& {0.8631}
& {34.12}
& {0.9479}
\\
ELAN~\cite{zhang2022elan} & $3\times$  &  629K & 90.1G
&34.61
&0.9288 
&30.55
&0.8463
&29.21 
&0.8081 
&28.69
&0.8624
&34.00
&0.9478
\\
SRformer-light~\cite{zhou2023srformer} & $3\times$  &  861K & 105.4G
&\second{34.67} 
&\second{0.9296} 
&\second{30.57}
&\second{0.8469}
&\second{29.26}
&\second{0.8099}
&\second{28.81}
&\second{0.8655}
&\second{34.19}
&\second{0.9489}
\\
\rowcolor[HTML]{EFEFEF} 
MambaIRv2-light & $3\times$  & 781K & 126.7G 
&\best{34.71}
&\best{0.9298}	
&\best{30.68}	
&\best{0.8483}	
&\best{29.26}	
&\best{0.8098}	
&\best{29.01}	
&\best{0.8689}	
&\best{34.41}	
&\best{0.9497}
\\
\midrule
CARN~\cite{ahn2018fast} & $4\times$  & 1,592K & 90.9G
& 32.13
& 0.8937
& 28.60
& 0.7806
& 27.58
& 0.7349
& 26.07 
& 0.7837
& {30.47}
& {0.9084}
\\
LatticeNet~\cite{luo2020latticenet} & $4\times$  & 777K & 43.6G
& {32.30}
& {0.8962}
& {28.68}
& {0.7830}
& {27.62}
& {0.7367}
& {26.25}
& {0.7873}
& -
& -
\\
SwinIR-light~\cite{liang2021swinir} & $4\times$  & 930K & {63.6G} %
& {32.44}
& {0.8976}
& {28.77}
& {0.7858}
& {27.69}
& {0.7406}
& {26.47}
& {0.7980}
& {30.92}
& {0.9151}
\\
MambaIR-light~\cite{guo2024mambair} & $4\times$  &  924K & 84.6G 
& {32.42}
& {0.8977}
& {28.74}
& {0.7847}
& {27.68}
& {0.7400}
& {26.52}
& {0.7983}
& {30.94}
& {0.9135}
\\
ELAN~\cite{zhang2022elan} & $4\times$  & 640K& 54.1G 
&32.43 
&0.8975 
&28.78 
&0.7858 
&27.69 
&0.7406 
&26.54 
&0.7982 
&30.92
&0.9150
\\
SRformer-light~\cite{zhou2023srformer} & $4\times$ & 873K &62.8G 
&\second{32.51} 
&\second{0.8988} 
&\second{28.82} 
&\second{0.7872} 
&\second{27.73}
&\second{0.7422} 
&\second{26.67}
&\second{0.8032}
&\second{31.17} 
&\second{0.9165}
\\
\rowcolor[HTML]{EFEFEF} 
MambaIRv2-light & $4\times$  &  790K & 75.6G 
&\best{32.51}	
&\best{0.8992}	
&\best{28.84}	
&\best{0.7878}
&\best{27.75}	
&\best{0.7426}	
&\best{26.82}	
&\best{0.8079}	
&\best{31.24}	
&\best{0.9182}
\\
\bottomrule
\end{tabular}%
}
\end{table*}

\begin{figure*}[!t]
    \centering
    \includegraphics[width=0.995\linewidth]{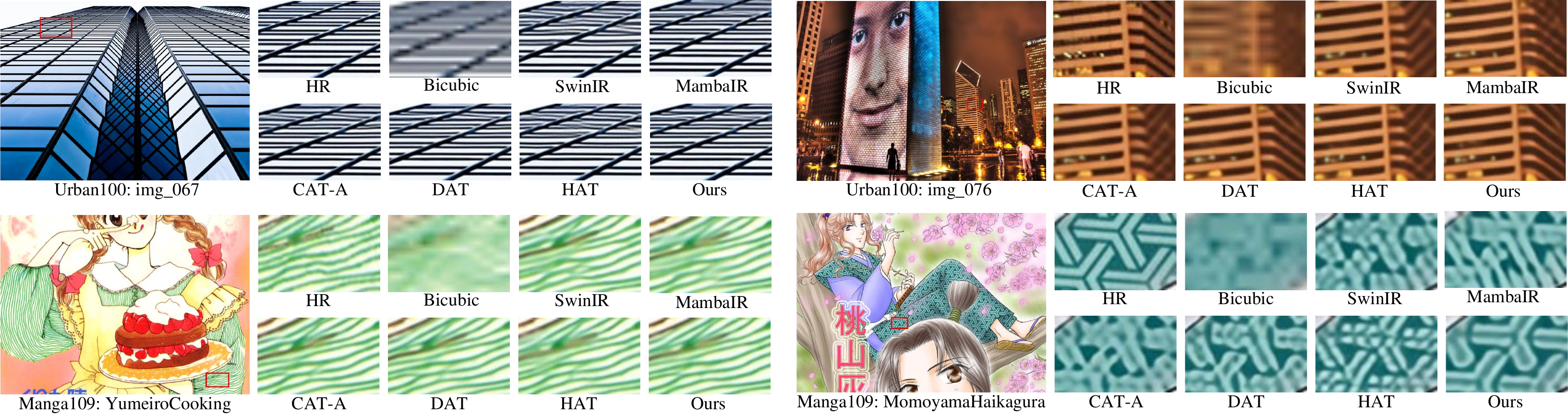}
    \vspace{-3mm}
    \caption{Qualitative comparison of our MambaIRv2 with different methods on $4\times$ classic image SR.}
    \vspace{-4mm}
    \label{fig:compareSR}
\end{figure*}

\noindent \textbf{Effectiveness of Different Components.}
As the core module of MambaIRv2, the Attentive State Space Module (ASSM), which contains the Attentive State-space Equation (ASE) and the Semantic Guided Neighboring (SGN), plays an important role in Mamba-based global modeling. In this ablation, we design three settings to verify the role of the different components. In the first setup, we directly remove the ASSM, leading to a pure Attention variant. In the second setup, we add the proposed ASSM but remove the SGN. The third setup corresponds to our proposed method. As shown in~\cref{tab:ablation-components}, only using window attention limits the receptive field to the local window, limiting the performance. Further, the addition of the ASE, which allows for the query of similar pixels across images, improves the performance by 0.05 dB/0.09 dB in Urban100/Manga109. Finally, after introducing SGN which can effectively overcome the long-range decay of Mamba, the final model achieves the best performance of 32.97/39.24 dB PSNR on Urban100/Manga109. The experiments validate the effectiveness of different components in the proposed method.

\noindent \textbf{Ablation on Attentive State-space Equation.}
In~\cref{sec:connect-attn-ssm}, we pointed out that the matrix $\mathbf{C}$ in the state-space equation in fact behaves similarly to the Query in the attention, following which we propose the ASE to insert global prompts into $\mathbf{C}$ by residual addition. In~\cref{tab:ablation-prompt-ASE}, we explore other inserting positions of the prompts in ASE. Since $\mathbf{B}$  can be analogized to the Key in attention, inserting the prompts into $\mathbf{B}$ also yields decent results, but slightly inferior to adding to $\mathbf{C}$. We attribute this to the fact that $\mathbf{C}$ is closer to the output end in the state-space equation~\cref{eq:ssm}, leading to a greater impact on the final result. Further, inserting prompts to the discrete time-step $\mathbf{\Delta}$ and the output $\mathbf{y}$ both fail to give satisfactory performance, which justifies our focus on the matrix $\mathbf{C}$ in the proposed ASE. Due to the page limit, we provide more ablation experiments in the \textit{Suppl.}.

\subsection{Comparison on Image Super-Resolution}

\begin{table*}[!t]
\centering
\caption{Quantitative comparison on \textit{\textbf{classic image super-resolution}} with state-of-the-art methods.}
\label{tab:classicSR}
\vspace{-3mm}
\setlength{\tabcolsep}{10pt}
\scalebox{0.74}{
\begin{tabular}{@{}l|c|c|cc|cc|cc|cc|cc@{}}
\toprule
 & & & \multicolumn{2}{c|}{\textbf{Set5}} &
  \multicolumn{2}{c|}{\textbf{Set14}} &
  \multicolumn{2}{c|}{\textbf{BSDS100}} &
  \multicolumn{2}{c|}{\textbf{Urban100}} &
  \multicolumn{2}{c}{\textbf{Manga109}} \\
\multirow{-2}{*}{Method} & \multirow{-2}{*}{scale} & \multirow{-2}{*}{\#param} & PSNR  & SSIM   & PSNR  & SSIM   & PSNR  & SSIM   & PSNR  & SSIM   & PSNR  & SSIM   \\ \midrule
EDSR~\cite{lim2017enhanced}   & $2\times$ & 42.6M & 38.11 & 0.9602 & 33.92 & 0.9195 & 32.32 & 0.9013 & 32.93 & 0.9351 & 39.10 & 0.9773 \\
RCAN~\cite{zhang2018image}   & $2\times$ &15.4M& 38.27 & 0.9614 & 34.12 & 0.9216 & 32.41 & 0.9027 & 33.34 & 0.9384 & 39.44 & 0.9786 \\
SAN~\cite{dai2019second}    & $2\times$ & 15.7M&38.31 & 0.9620 & 34.07 & 0.9213 & 32.42 & 0.9028 & 33.10 & 0.9370 & 39.32 & 0.9792 \\
IPT~\cite{chen2021pre} & $2\times$ &115M& 38.37 & - &34.43 &-& 32.48&-& 33.76& -& -& - \\
SwinIR~\cite{liang2021swinir} & $2\times$ &11.8M& 38.42 & 0.9623 & 34.46 & 0.9250 & 32.53 & 0.9041 & 33.81 & 0.9427 & 39.92 & 0.9797 \\
EDT~\cite{li2021edt}  & $2\times$ & 11.5M& 38.45 & 0.9624 & 34.57 &0.9258& 32.52&0.9041& 33.80& 0.9425& 39.93& 0.9800 \\
MambaIR~\cite{guo2024mambair} & $2\times$ & 20.4M& 38.57&0.9627&34.67&0.9261&32.58 &0.9048 &34.15& 0.9446 & 40.28 &0.9806\\
CAT-A~\cite{chen2022cat}  & $2\times$ & 16.5M& 38.51 &0.9626&34.78&0.9265&32.59 &0.9047 &34.26&0.9440 &40.10 &0.9805\\
DAT~\cite{chen2023dual}  & $2\times$ & 14.8M& 38.58 &0.9629&34.81&0.9297&32.61 &0.9051 &34.37&0.9458 &40.33 &0.9807\\
HAT~\cite{chen2023activating}  & $2\times$ &  20.6M&38.63 &0.9630&34.86&0.9274&32.62 &{0.9053} &34.45&0.9466 &40.26 &0.9809\\
\rowcolor[HTML]{EFEFEF} 
MambaIRv2-S & $2\times$ &9.6M& 38.53	&0.9627&	34.62&	0.9256&	32.59	&0.9048	&34.24	&0.9454	&40.27&0.9808 \\
\rowcolor[HTML]{EFEFEF} 
MambaIRv2-B & $2\times$& 22.9M& \second{38.65}&	\second{0.9631}	&\second{34.89}&\second{0.9275}&\second{32.62}&\second{0.9053}&\second{34.49}&\second{0.9468}&	\second{40.42}&\best{0.9810}
\\
\rowcolor[HTML]{EFEFEF} 
MambaIRv2-L & $2\times$&34.2M &\best{38.65}&	\best{0.9632}&	\best{34.93}	&\best{0.9276}&	\best{32.62}&	\best{0.9053}&	\best{34.60}&	\best{0.9475}&	\best{40.55}&	\second{0.9807} \\
\midrule
EDSR~\cite{lim2017enhanced} & $3\times$ & 42.6M&
34.65 & 0.9280 & 30.52 & 0.8462 & 29.25 & 0.8093 & 28.80 & 0.8653 & 34.17 & 0.9476\\
RCAN~\cite{zhang2018image} & $3\times$ & 15.4M&
34.74 & 0.9299 & 30.65 & 0.8482 & 29.32 & 0.8111 & 29.09 & 0.8702 & 34.44 & 0.9499\\
SAN~\cite{dai2019second} & $3\times$ & 15.7M&
34.75 & 0.9300 & 30.59 & 0.8476 & 29.33 & 0.8112 & 28.93 & 0.8671 & 34.30 & 0.9494\\
IPT~\cite{chen2021pre} & $3\times$ & 115M&
34.81 & - & 30.85 & - & 29.38 & - & 29.49 & - & - & - \\
SwinIR~\cite{liang2021swinir} & $3\times$ & 11.8M&
34.97 & 0.9318 & 30.93 & 0.8534 & 29.46 & 0.8145 & 29.75 & 0.8826 & 35.12 & 0.9537\\
EDT~\cite{li2021edt} & $3\times$ & 11.5M&
34.97 & 0.9316 & 30.89 & 0.8527 & 29.44 & 0.8142 & 29.72 & 0.8814 & 35.13 & 0.9534\\
MambaIR~\cite{guo2024mambair} & $3\times$ & 20.4M&
35.08 & 0.9323 &30.99 & 0.8536 & 29.51 & 0.8157 & 29.93 & 0.8841 & 35.43 & 0.9546\\
CAT-A~\cite{chen2022cat} & $3\times$ & 16.5M&
35.06 & 0.9326 & 31.04 & 0.8538 & 29.52 &0.8160 & 30.12 & 0.8862 & 35.80 &0.9546\\
DAT~\cite{chen2023dual} & $3\times$ & 14.8M&
35.16 & 0.9331 & 31.11 & 0.8550 & 29.55 & {0.8169} & 30.18 & 0.8886 & 35.59 &0.9554\\
HAT~\cite{chen2023activating} & $3\times$ & 20.6M&
35.07 & 0.9329 & 31.08 & 0.8555 & 29.54 & 0.8167 & 30.23 & 0.8896 & 35.53 &0.9552\\
\rowcolor[HTML]{EFEFEF} 
MambaIRv2-S & $3\times$ & 9.8M&35.09&0.9326&	31.07	&0.8547&	29.51&	0.8157&	30.08	&0.8871&	35.44&	0.9549\\
\rowcolor[HTML]{EFEFEF} 
MambaIRv2-B & $3\times$ & 23.1M&
\second{35.18} & \second{0.9334} & \second{31.12} & \second{0.8557} & \second{29.55} & \second{0.8169} & \second{30.28} & \second{0.8905} & \second{35.61} & \second{0.9556}\\
\rowcolor[HTML]{EFEFEF} 
MambaIRv2-L & $3\times$ & 34.2M&
\best{35.16} & \best{0.9334} & \best{31.18} & \best{0.8564} & \best{29.57} & \best{0.8175} & \best{30.34} & \best{0.8912} & \best{35.72} &\best{0.9561}\\
\midrule
EDSR~\cite{lim2017enhanced}   & $4\times$ & 43.0M & 32.46 & 0.8968 & 28.80 & 0.7876 & 27.71 & 0.7420 & 26.64 & 0.8033 & 31.02 & 0.9148 \\
RCAN~\cite{zhang2018image}   & $4\times$ & 15.6M &32.63 & 0.9002 & 28.87 & 0.7889 & 27.77 & 0.7436 & 26.82 & 0.8087 & 31.22 & 0.9173 \\
SAN~\cite{dai2019second}      & $4\times$ &15.7M& 32.64 & 0.9003 & 28.92 & 0.7888 & 27.78 & 0.7436 & 26.79 & 0.8068 & 31.18 & 0.9169 \\
IPT ~\cite{chen2021pre}          & $4\times$ & 116M& 32.64 &-& 29.01 &- &27.82 &-& 27.26& -& -& - \\
SwinIR~\cite{liang2021swinir}    & $4\times$ & 11.9M& 32.92 & 0.9044 & 29.09 & 0.7950 & 27.92 & 0.7489 & 27.45 & 0.8254 & 32.03 & 0.9260 \\
EDT~\cite{li2021edt}   & $4\times$ & 11.6M&  32.82 &0.9031& 29.09& 0.7939& 27.91& 0.7483 &27.46& 0.8246 &32.05& 0.9254 \\
MambaIR~\cite{guo2024mambair}   & $4\times$ &20.4M&33.03&0.9046&29.20&0.7961&{27.98}&{0.7503}&{27.68}&0.8287&{32.32}&{0.9272} \\
CAT-A~\cite{chen2022cat}    & $4\times$ & 16.6M &33.08 &0.9052& 29.18& 0.7960& 27.99& 0.7510& 27.89& 0.8339& 32.39& 0.9285\\
DAT~\cite{chen2023dual}    & $4\times$ & 14.8M&  33.08 &0.9055& 29.23 &0.7973& 28.00& 0.7515& 27.87& 0.8343& 32.51& 0.9291\\
HAT~\cite{chen2023activating}    & $4\times$ & 20.8M &33.04 &0.9056& 29.23& 0.7973& \second{28.00}& \second{0.7517}& \second{27.97}& \second{0.8368}& 32.48&0.9292 \\
\rowcolor[HTML]{EFEFEF} 
MambaIRv2-S    & $4\times$ & 9.8M& 32.99	&0.9037	&{29.23}	&0.7965&	27.97&	0.7502	&27.73&	0.8307	&32.33&	0.9276\\
\rowcolor[HTML]{EFEFEF} 
MambaIRv2-B  & $4\times$ & 23.1M& \second{33.14} & \second{0.9057} & \second{29.23} & \second{0.7975} & 28.00 & 0.7511 & {27.89} & {0.8344} & \second{32.57} & \second{0.9295} \\
\rowcolor[HTML]{EFEFEF} 
MambaIRv2-L    & $4\times$ & {34.2M}& \best{33.19} & \best{0.9062} & \best{29.29} & \best{0.7982} & \best{28.01} & \best{0.7521} & \best{28.07} & \best{0.8383} & \best{32.66} & \best{0.9304} \\
\bottomrule
\end{tabular}%
}
\vspace{-4mm}
\end{table*}

\noindent \textbf{Lightweight Image Super-Resolution.} 
Following previous works~\cite{zhou2023srformer,luo2020latticenet}, we also report the number of parameters (\#param) and MACs (upscaling a low-resolution image to $1280 \times 720$ resolution) as the efficiency metric. The results in~\cref{tab:lightSR} show that our \NAME outperforms the state-of-the-art method SRFormer~\cite{zhou2023srformer} using even significantly fewer parameters. For instance, our MambaIRv2-light outperforms SRFormer-light by 0.35dB on $2\times$ Urban100 with 79K fewer \#param. This experiment validates the efficiency and effectiveness of the proposed method.

\begin{table}[!tb]
\centering
\caption{
Complexity comparison with state-of-the-art methods on the $2 \times$ classic SR with output size $256 \times 256$.
}
\label{tab:complexity-compare}
\setlength{\tabcolsep}{4pt}
\vspace{-3mm}
\scalebox{0.8}{
\begin{tabular}{l|c|c|cccc}
\toprule
\multirow{2}{*}{Models} & \multirow{2}{*}{\#param} & \multirow{2}{*}{MACs} & \multicolumn{2}{c}{\textbf{Urban100}} & \multicolumn{2}{c}{\textbf{Manga109}} \\
            &       &      & PSNR  & SSIM   & PSNR  & SSIM   \\ \midrule
CAT-A~\cite{chen2022cat} & 16.6M & 350.7G & 34.26 & 0.9440 & 40.10 & 0.9805 \\
DAT~\cite{chen2023dual}  & 14.8M & 265.7G & 34.37 & 0.9458 & 40.33 & 0.9807 \\
HAT~\cite{chen2023activating}   & 20.8M & 514.9G & 34.45 & 0.9466 & 40.26 & 0.9809 \\
\rowcolor[HTML]{EFEFEF} 
MambaIRv2-S & 9.6M  & 192.9G  & 34.24 & 0.9454 & 40.27 & 0.9808 \\
\rowcolor[HTML]{EFEFEF} 
MambaIRv2-B & 22.9M & 445.8G & 34.49 & 0.9468 & 40.42 & 0.9810 \\
\rowcolor[HTML]{EFEFEF} 
MambaIRv2-L & 34.2M & 664.5G & 34.60 & 0.9475 & 40.55 & 0.9807 \\ \bottomrule
\end{tabular}%
}
\vspace{-6mm}
\end{table}

\noindent \textbf{Classic Image Super-Resolution.}
\cref{tab:classicSR} gives the comparison results of our MambaIRv2 with different model sizes to existing state-of-the-art classic SR methods. Thanks to the attentive state space modeling, our proposed method achieves the best performance across most five benchmark datasets and up-sample scales. For example, our MambaIRv2-B exceeds HAT~\cite{chen2023activating} by 0.16dB on the $2\times$ Manga109 dataset. Interestingly, even the MambaIRv2-S with 9.6M \#param, outperforms the previous 20.4M MambaIR~\cite{guo2024mambair} by 0.09dB PSNR on the $2\times$ Urban100 dataset, further demonstrating our MambaIRv2 serves as an elegant balance of performance and efficiency. Finally, our method also shows promising scaling-up capabilities. When we scale up the \#param to 34.2M to obtain the MambaIRv2-L model, this larger model achieved steady performance gains compared to its base counterpart, \textit{e.g.}, 0.18dB PSNR gains on $4\times$ Urban100. The visual comparisons are shown in~\cref{fig:compareSR}, and our method can facilitate the reconstruction of sharp edges and natural textures.

\noindent \textbf{Model Complexity Comparison.}
As shown in~\cref{tab:complexity-compare}, our MambaIRv2-S model, which uses only 55.0\% of the MACs compared to CAT-A~\cite{chen2022cat}, outperforms CAT-A by 0.17dB PSNR on Manga109. Additionally, our MambaIRv2-B model, which roughly matches the \#param of HAT~\cite{chen2023activating}, achieves a 13.4\% reduction in MACs, while delivering 0.04/0.16dB PSNR improvements on Urban100/Manga109. The above results demonstrate that our MambaIRv2 strikes a sweet spot between performance and efficiency.

\begin{table*}[tb]
\centering
\caption{Quantitative comparison on \textit{\textbf{JPEG compression artifact reduction}} under different quality factors $q$.}
\label{tab:jpeg-car}
\vspace{-3mm}
\setlength{\tabcolsep}{9pt}
\scalebox{0.76}{
\begin{tabular}{l|c|cc|cc|cc|cc|cc|cc}
\toprule
\multirow{2}{*}{Dataset} & \multirow{2}{*}{$q$} &  \multicolumn{2}{c|}{RNAN~\cite{RNAN}} & \multicolumn{2}{c|}{RDN~\cite{RDN}} & \multicolumn{2}{c|}{DRUNet~\cite{DRUNet}} & \multicolumn{2}{c|}{SwinIR~\cite{liang2021swinir}} & 
\multicolumn{2}{c|}{MambaIR~\cite{guo2024mambair}} &
\multicolumn{2}{c}{Ours} \\
& & PSNR & SSIM & PSNR & SSIM & PSNR & SSIM & PSNR & SSIM & PSNR & SSIM & PSNR & SSIM	\\ \hline	
\multirow{3}{*}{Classic5}	
& 10 
& 29.96 & 0.8178 
& 30.00 & 0.8188 
& 30.16 & 0.8234 
& 30.27 & 0.8249	
& \second{30.27} & \second{0.8256}	
& \best{30.37}	& \best{0.8269}
\\
& 30 
& 33.38 & 0.8924 
& 33.43 & 0.8930 
& 33.59 & 0.8949 
& 33.73 & 0.8961 
&\second{33.74} & \second{0.8965} 
&\best{33.81}	& \best{0.8970}
\\
& 40
& 34.27 & 0.9061 
& 34.27 & 0.9061 
& 34.41 & 0.9075
& 34.52 & 0.9082 
& \second{34.53} & \second{0.9084}
& \best{34.64} & \best{0.9093}
\\ \hline
\multirow{3}{*}{LIVE1}
& 10
& 29.63 & 0.8239
& 29.67 & 0.8247
& 29.79 & 0.8278
& 29.86 & 0.8287
& \second{29.88} & \second{0.8301}
& \best{29.91}	& \best{0.8301}
\\
& 30
& 33.45 & 0.9149
& 33.51 & 0.9153
& 33.59 & 0.9166
& 33.69 & 0.9174
& \second{33.72} & \second{0.9179}
& \best{33.73}	&\best{0.9179}
\\
& 40
& 34.47 & 0.9299
& 34.51 & 0.9302
& 34.58 & 0.9312
& 34.67 & 0.9317
& \second{34.70} & \second{0.9320}
& \best{34.73} &	\best{0.9323}
\\ \bottomrule   
\end{tabular}
}
\end{table*}

\begin{table}[!tb]
\centering
\caption{Quantitative comparison of PSNR on \textit{\textbf{gaussian color image denoising}} $\sigma=15$ with state-of-the-art methods.}
\label{tab:guassian-denoise}
\vspace{-3mm}
\setlength{\tabcolsep}{3.8pt}
\scalebox{0.9}{
\begin{tabular}{@{}l|cccc@{}}
\toprule
Method    & CBSD68 & Kodak24 & McMaster & Urban100 \\ \midrule
IRCNN~\cite{IRCNN}    & 33.86  & 34.69   & 34.58    & 33.78    \\
FFDNet~\cite{FFDNet}      & 33.87  & 34.63   & 34.66    & 33.83    \\
DnCNN~\cite{DnCNN}      & 33.90  & 34.60   & 33.45    & 32.98    \\
DRUNet~\cite{DRUNet}   & 34.30  & 35.31   & 35.40    & 34.81    \\
SwinIR~\cite{liang2021swinir}     & 34.42  & 35.34   & 35.61    & 35.13    \\
Restormer~\cite{zamir2022restormer} & 34.40  & 35.35   & 35.61    & 35.13    \\
Xformer~\cite{zhang2023xformer}   & 34.43  & 35.39   & 35.68    & 35.29    \\
MambaIR~\cite{guo2024mambair} & \second{34.48}  & \second{35.42}   & \second{35.70}    & \second{35.37}    \\
\rowcolor[HTML]{EFEFEF} 
MambaIRv2 & \best{34.48}  & \best{35.43}   & \best{35.73}    & \best{35.42}    \\ \bottomrule
\end{tabular}%
}
\end{table}

\begin{table}[!tb]
\centering
\caption{Comparison to MambaIR with different scanning modes on $2\times$ lightweight SR. The ``MambaIR-n'' indicates the MambaIR~\cite{guo2024mambair} variant with $n$ scanning directions.}
\vspace{-3mm}
\label{tab:discussion-direction-efficiency}
\setlength{\tabcolsep}{1.8pt}
\scalebox{0.88}{
\begin{tabular}{@{}l|cc|cccc@{}}
\toprule
\multirow{2}{*}{settings} & \multirow{2}{*}{\#param} & \multirow{2}{*}{MACs} & \multicolumn{2}{c}{\textbf{Urban100}} & \multicolumn{2}{c}{\textbf{Manga109}} \\
          &       &      & PSNR & SSIM & PSNR & SSIM \\ \midrule
MambaIR-1~\cite{guo2024mambair} & 987K  & 291G &     32.82 & 0.9348     & 39.19     &  0.9776     \\
MambaIR-2~\cite{guo2024mambair} & 1.11M & 383G &  32.86    &  0.9450     &  39.26 &   0.9778   \\
MambaIR-4~\cite{guo2024mambair} & 1.36M & 568G &   32.92 &  0.9356    & 39.31   &   0.9779   \\
\rowcolor[HTML]{EFEFEF} 
MambaIRv2 &   \textbf{774K}    &  \textbf{286G}  &   \textbf{33.26}  &  \textbf{0.9378}   &   \textbf{39.39}   & \textbf{0.9786}    \\ \bottomrule
\end{tabular}%
}
\vspace{-2mm}
\end{table}

\subsection{Comparison on JPEG CAR}
\cref{tab:jpeg-car} gives the results on JPEG compression reduction. It can be seen that the proposed \NAME achieves the best performance on all testing datasets across all quality factors. For example, our \NAME suppresses MambaIR~\cite{guo2024mambair} by 0.11dB PSNR with $q=40$ on the Classic5 dataset, demonstrating the effectiveness of our \NAME on other restoration tasks.

\subsection{Comparison on Image Denoising}
We further include the gaussian color image denoising task for further validation. Notably, to maintain architectural consistency across different restoration tasks, we retain the straight-through structure and avoid using the UNet architecture, which has been shown to be particularly advantageous for denoising tasks~\cite{chen2023xrestormer}. The results presented in~\cref{tab:guassian-denoise} demonstrate that MambaIRv2 outperforms all other models across the datasets. In particular, it surpasses U-shaped Restormer~\cite{zamir2022restormer} by even 0.29dB PSNR  on the Urban100 dataset. This experiment validates our MambaIRv2 serves as a generalized image restoration backbone.

\subsection{Discussion}

\noindent \textbf{Benefits from Reduced Scan Directions.}
Compared to the previous MambaIR~\cite{guo2024mambair}, which performs 4 scans in pixel space, a significant advantage of our MambaIRv2 is that it requires only a single scan in the semantic space. As shown in~\cref{tab:discussion-direction-efficiency}, our MambaIRv2 is not only efficient but also boosts performance. For example, MambaIRv2 reduces even \textbf{43\%} of the number of parameters and \textbf{50\%} computational burden compared to the standard MambaIR, while still suppresses by \textbf{0.34dB} PSNR on $2\times$ Urban100.

\begin{figure}[!t]
    \centering
    \includegraphics[width=1\linewidth]{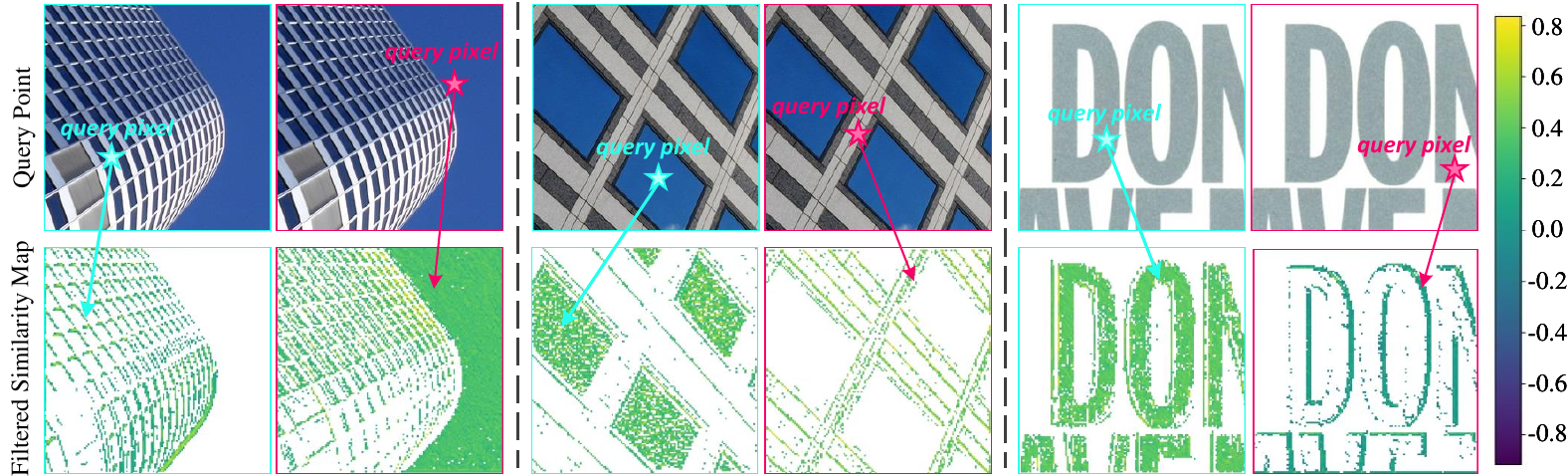}
    \vspace{-6mm}
    \caption{The visualization of the attentive state space. We compute the cosine similarity between the prompt corresponding to the query pixel and the matrix $\mathbf{C}$. We filter out low-similarity points for presentation clarity. More examples are provided in the \textit{Suppl.}.}
    \vspace{-3mm}
    \label{fig:attntive-map}
\end{figure}

\noindent \textbf{Visualization of Attentive State Space.}
In the proposed attentive state space equation,  the prompts play an important role in representing similar pixels across the whole image to facilitate the query pixel seeing out of the scanned sequence. As shown in ~\cref{fig:attntive-map}, it can be seen that the query pixel is empowered to attend to its corresponding semantic part in the image through the prompt, thus enabling global information aggregation similar to attention mechanism.

\section{Conclusion}

In this work, we introduce MambaIRv2 to enhance state-space restoration models by addressing the causal modeling nature of Mamba. We propose the attentive state-space equation that incorporates prompt learning for enlarged token perception as well as scanning only once. Additionally, we introduce semantic guided neighboring which positions similar pixels closer to handle the long-range decay. These innovations enable  MambaIRv2 to integrate ViT-like non-causal abilities into Mamba-based models to implement the attentive state space restoration. Extensive experiments confirm our MambaIRv2 as an efficient, high-performing backbone for image restoration.

{
    \small{
    \bibliographystyle{ieeenat_fullname}
    \bibliography{main}}
}

\clearpage
\maketitlesupplementary
\appendix
\renewcommand{\theequation}{A.\arabic{equation}}
\setcounter{equation}{0}
\renewcommand{\thetable}{A.\arabic{table}}
\setcounter{table}{0}
\renewcommand{\thefigure}{A.\arabic{figure}}
\setcounter{figure}{0}

\section{Efficiency Comparison on Large Inputs}

Benefiting from the Mamba architecture, our proposed MambaIRv2 can achieve global pixel utilization. However, as an inevitable side effect, the global receptive field is usually accompanied by an increased computational cost since the model needs to process more tokens at once. Therefore, it is necessary to validate the efficiency on large resolution images. Here, we point out that, benefiting  from our single-directional scan, our MambaIRv2 can in fact achieve a similar computational cost as the advanced Swin-Transformer~\cite{liu2021swin} based method HAT~\cite{chen2023activating}. In~\cref{tab:suppl-macs-resolution}, we give the MACs of our MambaIRv2 and HAT under varying input resolutions. As one can see, our MambaIRv2-B, which has a roughly similar number of parameters as HAT~\cite{chen2023activating}, is more efficient than HAT from resolution $64\times 64$ to $1024 \times 1024$. For example, on the $256 \times 256$ resolution, which is a common inference patch size, our method achieves a 30\% savings in computational cost metric MACs. At the high-resolution setting of $1024\times1024$, our method achieves fewer MACs than HAT. It is worth noting that in addition to this impressive efficiency, our MambaIRv2 still outperforms HAT by a noticeable margin, which has been extensively verified in the main paper.

\section{More Ablation Results}

\subsection{Ablation on Prompt Learning}

In the proposed ASE, we use learnable prompts to absorb information of similar pixels across the whole image, which will be later inserted into the state space modeling to help the query pixel to see the unscanned tokens. The proposed prompt learning contains two key hyperparameters, namely the size of the prompt pool $T$, and the internal rank $r$ in the semantic decoupling. In this section, we perform hyperparameter ablation to investigate the impact of different $T$ and $r$ on the performance. As shown in~\cref{tab:suppl-hyper-prompt}, when the $r$ is small, increasing the number of prompts $T$ can steadily improve performance. For example, when $r$ = 16, increasing $T$ from 64 to 128 can result in a 0.03dB improvement on Manga109. However, when $r$ is large, increasing the size of the prompt pool sometimes instead results in a slight performance drop. A similar observation also appears in the inner-rank $r$. In practice, we choose a moderate $T\times r = 32 \times 64$ considering the performance and efficiency trade-off.

\begin{table}[!tb]
\centering
\caption{The computational cost MACs with images of different resolutions. We compare our MambaIRv2-B and HAT~\cite{chen2023activating}. We adopt the $4\times$ classical SR task and set the output size from $64\times$ to $1024\times1024$.}
\label{tab:suppl-macs-resolution}
\setlength{\tabcolsep}{1.5pt}
\scalebox{0.83}{
\begin{tabular}{@{}l|ccccc@{}}
\toprule
models      & $64\times 64$ & $128 \times 128$ & $256 \times 256$ & $512 \times 512$ & $1024 \times 1024$ \\ \midrule
HAT~\cite{chen2023activating}  & 26.05G        & 58.62G           & 162.85G          & 527.63G          & 1882.28G           \\
\rowcolor[HTML]{EFEFEF} 
MambaIRv2 & 7.12G         & 28.49G           & 113.97G          & 455.89G           & 1823.04G           \\ \bottomrule
\end{tabular}%
}
\end{table}

\begin{table}[!t]
\centering
\caption{Ablation experiments on the hyper-parameters of the number of prompts $T$ in the prompt pool, and the inner rank $r$ in the semantic decoupling.}
\label{tab:suppl-hyper-prompt}
\setlength{\tabcolsep}{6pt}
\scalebox{0.8}{
\begin{tabular}{@{}l|cccccc@{}}
\toprule
\multirow{2}{*}{$r\times T$} &\multicolumn{2}{c}{\textbf{Set14}}& \multicolumn{2}{c}{\textbf{Urban100}} & \multicolumn{2}{c}{\textbf{Manga109}} \\
            & PSNR & SSIM & PSNR & SSIM& PSNR & SSIM \\ \midrule
16 $\times$ 64 & 33.90&0.9205& 32.96	& 0.9359&39.19&0.9781    \\
16 $\times$ 128 & 33.91&0.9205& 32.96&	0.9354&	39.22&0.9783    \\
32 $\times$ 128 & \textbf{33.97}&0.9210& \textbf{32.97}& \textbf{0.9360}&39.20&	0.9783\\
\rowcolor[HTML]{EFEFEF} 
32 $\times$ 64 & 33.95&	\textbf{0.9213} & \textbf{32.97}&	0.9355&	\textbf{39.24} &	\textbf{0.9784}   \\ \bottomrule
\end{tabular}%
}
\end{table}

\subsection{Visualization of Semantic Neighboring}

\begin{figure}[!t]
    \centering
    \includegraphics[width=1\linewidth]{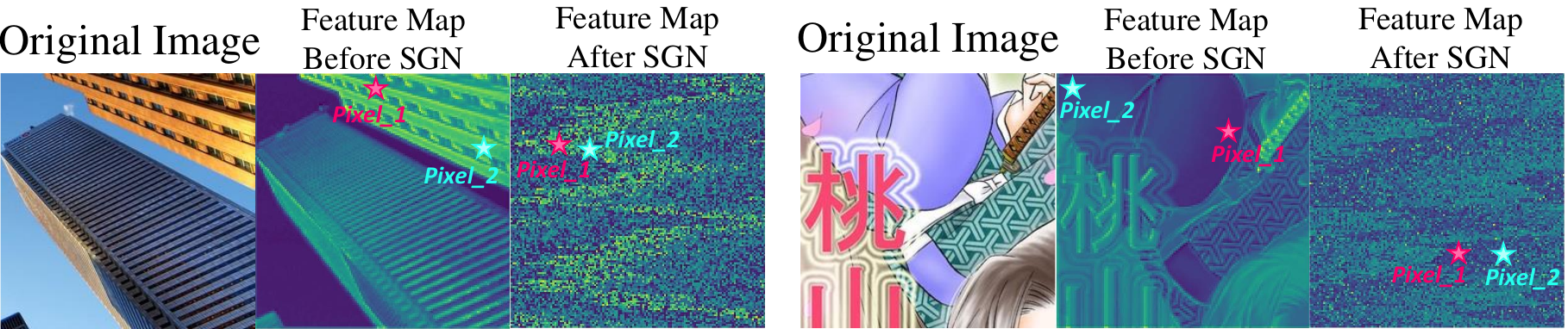}
    \caption{Visualization of the effectiveness of the proposed SGN. Before SGN, the semantically similar pixels belonging to the same object in the original feature map are far apart. After SGN, these pixels are spatially close to each other, thus facilitating strong interactions in the unfolded 1D sequence.
    }
    \label{fig:suppl-visualization-sgn-unflod}
\end{figure}

In the proposed Semantic Guided Neighboring (SGN), we restructure the image so that semantically similar pixels are also spatially close in the unfolded 1D sequence. In this section, we visualize the learned restructured image in~\cref{fig:suppl-visualization-sgn-unflod} for better understanding. It can be seen that the previous distant pixels with similar semantics in the original feature map become spatially close after the restructuring of the  SGN. By placing semantically similar pixels closer, the proposed SGN alleviates Mamba's long-range decay problem resulting from the causal modeling nature and thus facilitates better exploit those distant but similar pixels.

\section{Comparison on Receptive Field}

\begin{figure*}[!t]
    \centering
    \includegraphics[width=0.7\linewidth]{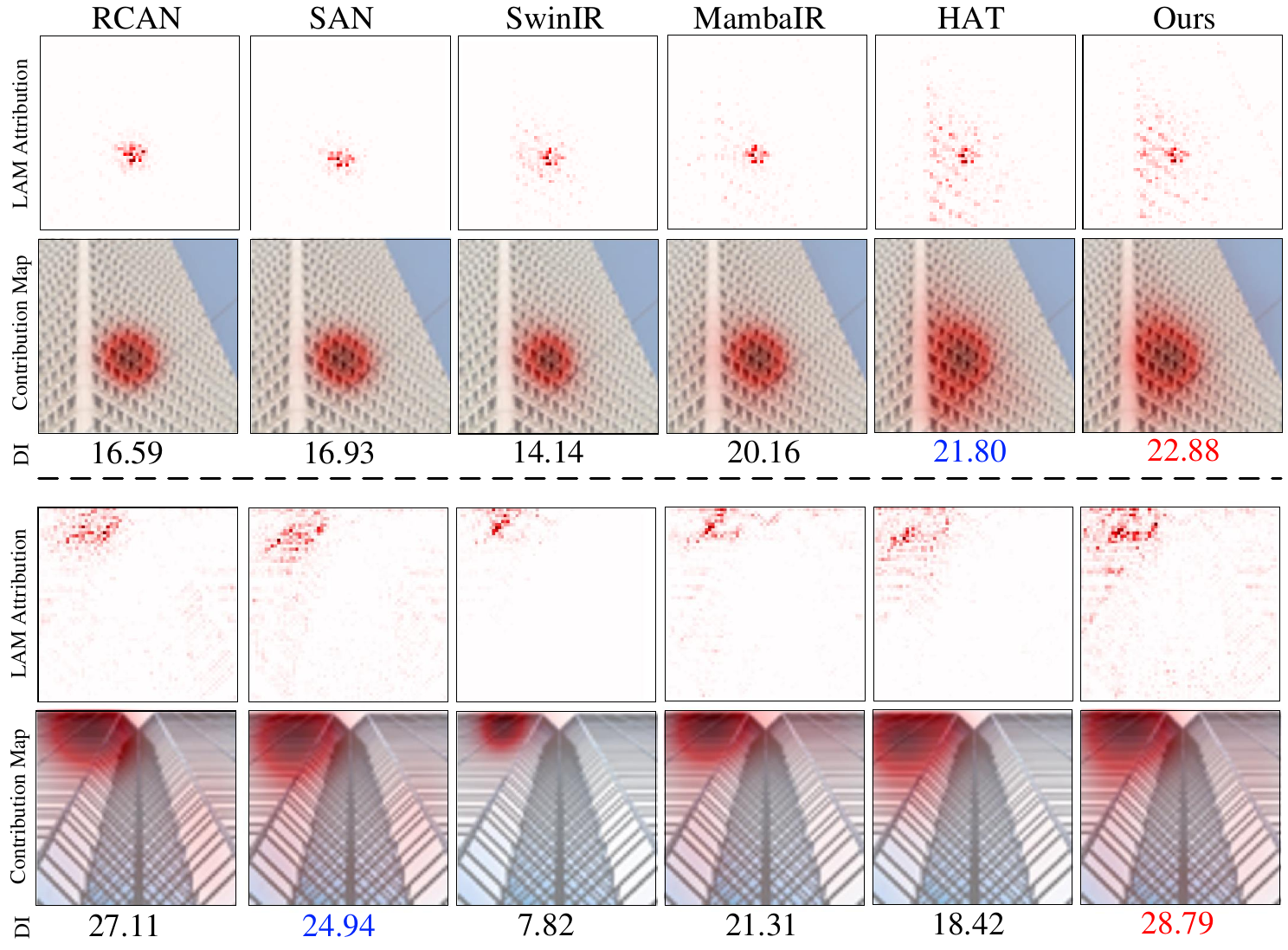}
    \caption{The LAM visualization~\cite{gu202lam} comparison with different methods The diffusion index reflects the range of involved pixels. A higher DI represents a wider range of utilized pixels.}
     \label{fig:suppl-lam-visualization}
\end{figure*}

\begin{figure*}[!t]
    \centering
    \includegraphics[width=0.8\linewidth]{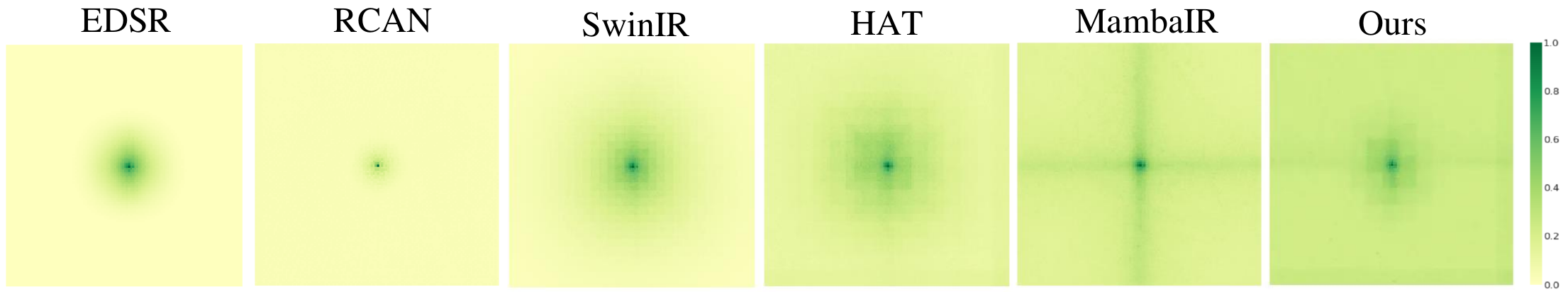}
    \caption{The Effective Receptive Field (ERF) visualization~\cite{luo2016understanding,ding2022scaling} for EDSR~\cite{lim2017enhanced}, RCAN~\cite{zhang2018image}, SwinIR~\cite{liang2021swinir}, HAT~\cite{chen2023activating}, MambaIR~\cite{guo2024mambair}, and the proposed \NAME. A larger ERF is indicated by a more extensively distributed dark area. The proposed \NAME achieves a significant global effective receptive field.}
    \label{fig:suppl-erf-visualization}
\end{figure*}

As pointed out in previous work~\cite{guo2024mambair}, a significant advantage of the Mamba architecture is the practical global receptive field, which helps the model activate more pixels to improve restoration performance. Here, we give visualization comparison results of LAM~\cite{gu202lam} and ERF~\cite{luo2016understanding} with other strong baselines. First, the \cref{fig:suppl-lam-visualization} gives the results of the LAM attribution map. One can see that our \NAME can activate more pixels than other state-of-the-art methods HAT~\cite{chen2023activating} by presenting a wider LAM attribution and a higher DI, thus resulting in higher-quality restoration results. Second, the ~\cref{fig:suppl-erf-visualization} further gives the effective receptive filed visual comparison with other methods. Our MambaIRv2 exhibits darker colors across the entire image, indicating the global perception of the proposed method.

It is noteworthy that the ERF visualization in~\cref{fig:suppl-erf-visualization} can also demonstrate the effectiveness of the proposed non-causal modeling strategy in our MambaIRv2. In detail, the ERF visualization of MambaIR~\cite{guo2024mambair} exhibits a clear crisscrossing, which is a clear sign of the causal modeling property as the center pixel can only utilizes its previous pixels in the scanned 1D sequence. In contrast, our proposed MambaIRv2, which aims at eliminating causal modeling in Mamba for image restoration, does not exhibit such unfavorable crisscrossing, demonstrating the validity of our proposed non-causal modeling.

\section{More Implementation Details}
We employ the DF2K~\cite{timofte2017ntire,lim2017enhanced} dataset to train models on classic SR and use DIV2K only to train lightweight SR models. Moreover, we use Set5~\cite{bevilacqua2012low}, Set14~\cite{zeyde2012single}, B100~\cite{martin2001database}, Urban100~\cite{huang2015single}, and Manga109~\cite{matsui2017sketch} to evaluate the effectiveness of different SR methods. For Gaussian color image denoising and JPEG CAR, we utilize DIV2K~\cite{timofte2017ntire}, Flickr2K~\cite{lim2017enhanced}, BSD500~\cite{arbelaez2010contour}, and WED~\cite{ma2016waterloo} as our training datasets. Our testing datasets for guassian color image denoising includes BSD68~\cite{martin2001database}, Kodak24~\cite{kao}, McMaster~\cite{zhang2011color}, and Urban100~\cite{huang2015single}. And we use  Classic5~\cite{foi2007classic5} and LIVE1~\cite{sheikh2005live} datasets to evaluate the performance of the JPEG CAR task. The performance is evaluated using PSNR and SSIM on the Y channel from the YCbCr color space.

\section{Comparison to ATD}
% difference from ATD reoder
The Adaptive Token Dictionary (ATD)~\cite{zhang2024transcending} which can generate input-specific tokens/prompts to help the query pixel see out of the window, appears close to our proposed MambaIRv2, with both including additional prompts for obtaining more information. Here, we summarize the main differences between them in the following aspects. First, the goals of introducing prompts in these two methods are clearly different. The ATD uses prompts to overcome the limited receptive field in the window attention, while our MambaIRv2 aims to mitigate the causal modeling of the Mamba. Second, the utilization of prompts for seeing beyond the scanned sequence in our MambaIRv2 is well-motivated. Specifically, we mathematically analyze the difference between state space and attention in the main paper, based on which we propose to add prompts to the $\mathbf{C}$ matrix in the state equation to attentively query relevant pixels across the image. In contrast, ATD adopts intuitive cross-attention to incorporate prompts into the feature map. Third, the way in which the prompts are generated is different. Specifically, ATD uses the attention map to implicitly obtain the category of each pixel. However, attention maps are even not available in Mamba, and thus we propose to design separate routing modules to explicitly learn the category of each pixel. In~\cref{tab:suppl-compare-atd}, we give the quantitative comparison of our MambaIRv2 against ATD, and it can be seen that our proposed method can achieve comparable performance to ATD. It should be noted that ATD~\cite{zhang2024transcending} is a highly optimized Transformer-based method since Transformer has been introduced to image restoration for many years. Given the Mamba-based methods are still in their infancy since the introduction of MambaIR~\cite{guo2024mambair}. It is promising for the Mamba-based method to achieve further performance improvements over its transformer counterparts.

\begin{table}[!tb]
\centering
\caption{Comparison with ATD~\cite{zhang2024transcending} on $2\times$ classic SR.}
\label{tab:suppl-compare-atd}
\setlength{\tabcolsep}{6pt}
\scalebox{0.9}{
\begin{tabular}{@{}l|ccccc@{}}
\toprule
methods & Set5  & Set14 & B100  & Urban100 & Manga109 \\ \midrule
ATD~\cite{zhang2024transcending}     & 38.61 & \textbf{34.95} & \textbf{32.65} & \textbf{34.70}    & 40.37    \\
\rowcolor[HTML]{EFEFEF} 
MambaIRv2    & \textbf{38.65} & 34.89 & 32.62 & 34.49    & \textbf{40.42}    \\ \bottomrule
\end{tabular}%
}
\end{table}

\section{Limitation and Future Works}

Our MamabIRv2 can effectively alleviate the inherent causal nature of Mamba architecture~\cite{gu2023mamba} benefiting from the proposed attentive state space modeling. Nonetheless, our work can be further improved in the future in the following aspects. First, the Mamba architecture emerges as the third backbone option for image restoration, in addition to CNNs and ViTs, which provide more solutions for designing image restoration networks. Therefore, an in-depth interpretability analysis about what exactly Mamba or ViT has learned during the restoration of an image is important for further understanding and network design. Second, although this work follows existing works~\cite{liang2021swinir} to cover multiple image restoration tasks, some other tasks such as image deblurring, dehazing and deraining can also be explored in the future. The implementation of the U-shaped MambaIRv2 backbone for these tasks to achieve further performance improvement is also interesting and promising~\cite{chen2023xrestormer}. Finally, despite the promising results shown, we would like to point out that the Mamba-based image restoration network is still in its early stages. With the increasing research interest in Mamba, it will be promising to study the state-space models for low-level vision.

\section{Proof for Long-range Decay}
% mmaba A^\alpha

As pointed out in the main paper, the causal property of Mamba leads to weak interactions between the query token and other remote tokens, \textit{i.e.}, the long-range decay. Here, given the condition of the causal modeling equation in Mamba, we attempt to derive the long-range decay as follows.

Formally, recall that the causal modeling of the state-space equation is given by:
\begin{equation}
\begin{aligned}
\label{eq:suppl-ssm}
&h_{i}=\mathbf{\overline{A}} h_{i-1}+\mathbf{\overline{{B}}} x_i,\\
&y_{i}=\mathbf{C} h_i+ \mathbf{D} x_i.
\end{aligned}
\end{equation}
Then, we can continuously iterate~\cref{eq:suppl-ssm} wit $i=0,1,\cdots k$. For example, setting $i=0$ turns~\cref{eq:suppl-ssm} into the following: 
\begin{equation}
\begin{aligned}
h_0&=\mathbf{\overline{{B}}}x_0\\
y_0&=\mathbf{C}h_0+\mathbf{D}x_0=\mathbf{C}\mathbf{\overline{{B}}}x_0+\mathbf{D}x_0
\end{aligned}
\end{equation}
After that, we can further set $i=1$ to obtain the following equation:
\begin{equation}
\begin{aligned}
h_1&=\mathbf{\overline{{A}}}h_0+\mathbf{\overline{{B}}}x_1=\mathbf{\overline{{A}}}\mathbf{\overline{{B}}}x_0+\mathbf{\overline{{B}}}x_1\\
y_1&=\mathbf{C}h_1+\mathbf{D}x_1=\mathbf{C}(\mathbf{\overline{{A}}}\mathbf{\overline{{B}}}x_0+\mathbf{\overline{{B}}}x_1)+\mathbf{D}x_1\\
&=\mathbf{C}\mathbf{\overline{{A}}}\mathbf{\overline{{B}}}x_0+\mathbf{C}\mathbf{\overline{{B}}}x_1 + \mathbf{D}x_1
\end{aligned}
\end{equation}
Set $i=2$ gives the following:
\begin{equation}
\begin{aligned}
h_2&=\mathbf{\overline{{A}}}h_1+\mathbf{\overline{{B}}}x_2=\mathbf{\overline{{A}}}(\mathbf{\overline{{A}}}\mathbf{\overline{{B}}}x_0+\mathbf{\overline{{B}}}x_1)+\mathbf{\overline{{B}}}x_2\\
&=\mathbf{\overline{{A}}}^2\mathbf{\overline{{B}}}x_0+\mathbf{\overline{{A}}}\mathbf{\overline{{B}}}x_1+\mathbf{\overline{{B}}}x_2 \\ 
y_2&=\mathbf{C}h_2+\mathbf{D}x_2 \\
&= \mathbf{C}(\mathbf{\overline{{A}}}^2\mathbf{\overline{{B}}}x_0+\mathbf{\overline{{A}}}\mathbf{\overline{{B}}}x_1+\mathbf{\overline{{B}}}x_2)+\mathbf{D}x_2\\
&=\mathbf{C}\mathbf{\overline{{A}}}^2\mathbf{\overline{{B}}}x_0+\mathbf{C}\mathbf{\overline{{A}}}\mathbf{\overline{{B}}}x_1+\mathbf{C}\mathbf{\overline{{B}}}x_2 + \mathbf{D}x_2
\end{aligned}
\end{equation}
By iterating continuously, we can generalize the output $y_k$ in the $k$-th time step being represented by $x_0 \cdots x_k$ as the following formula:
\begin{equation}
\label{eq:suppl-long-rang-decay}
y_k=\mathbf{C}\mathbf{\overline{{A}}}^k\mathbf{\overline{{B}}}x_0+\mathbf{C}\mathbf{\overline{{A}}}^{k-1}\mathbf{\overline{{B}}}x_1+\cdots+\mathbf{C}\mathbf{\overline{{B}}}x_k + \mathbf{D} x_k
\end{equation}
\cref{eq:suppl-long-rang-decay} actually quantitative the interaction between the $k$-th query token $x_k$ and all its previous $k$ tokens $x_0, x_1, \cdots, x_{k-1}$ in the causally scanned sequences to produce the $k$-th output $y_k$ of state-space model.
It can be clearly seen in~\cref{eq:suppl-ssm} that the contribution of $x_0$ to the generation of $y_k$ is weighted by $\mathbf{C}\mathbf{\overline{{A}}}^k\mathbf{\overline{{B}}}$, which is proportional to $\mathbf{\overline{{A}}}^k$. Since in the main paper we have empirically observed that the mean value of $\mathbf{\overline{{A}}}$ is statistically less than 1, as a result, when $k$ is large, \textit{i.e.}, when the two pixels are distant, the contribution of $x_0$ to $x_k$ is small, \textit{i.e.}, exhibiting long-range decay. If $x_0$ is very helpful to $x_k$, this decay can catastrophically impair the restoration of the $x_k$.

\begin{figure*}[!t]
    \centering
    \includegraphics[width=0.95\linewidth]{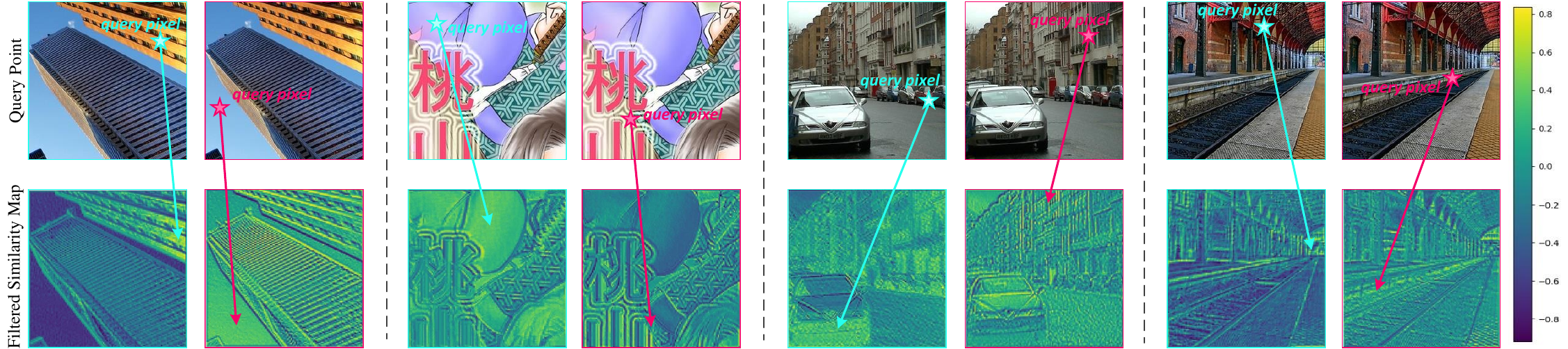}
    \caption{More visualization results on the attentive state space modeling.}
    \label{fig:suppl-more-attentive map.}
\end{figure*}

\end{document}